\documentclass[aps,prb,floatfix,reprint]{revtex4-1}

\usepackage{graphicx}
\usepackage{subfigure}
\usepackage{amsfonts}
\usepackage{amsmath}
\usepackage{amssymb}

\usepackage{color,soul}

\begin{document}

\title{Geometric rectification for nanoscale vibrational energy harvesting.}

\author{Ra\'ul A. Bustos-Mar\'un}
\thanks{\texttt{rbustos@famaf.unc.edu.ar}}
\affiliation{Instituto de F\'{\i}sica Enrique Gaviola (CONICET-UNC), Facultad de Matem\'{a}tica 
Astronom\'{\i}a, F\'{\i}sica y Computaci\'on and Facultad de Ciencias Qu\'{\i}micas, Universidad Nacional de C\'{o}rdoba,
Ciudad Universitaria, C\'{o}rdoba, 5000, Argentina.}

\begin{abstract}
In this work, we present a mechanism that, based on quantum-mechanical principles, allows one to recover kinetic energy at the nanoscale.
Our premise is that very small mechanical excitations, such as those arising from sound waves propagating through a nanoscale system or similar phenomena, can be quite generally converted into useful electrical work by applying the same principles behind conventional adiabatic quantum pumping.
The proposal is potentially useful for nanoscale vibrational energy harvesting where it can have several advantages.
The most important one is that it avoids the use of classical rectification mechanisms as it is based on what we call geometric rectification. 
We show that this geometric rectification results from applying appropriate but quite general initial conditions to damped harmonic systems coupled to electronic reservoirs.
We analyze an analytically solvable example consisting of a wire suspended over permanent charges where we find the condition for maximizing the pumped charge. We also studied the effects of coupling the system to a capacitor including the effect of current-induced forces and analyzing the steady-state voltage of operation. Finally, we show how quantum effects can be used to boost the performance of the proposed device.
\end{abstract}
\maketitle

\section{Introduction}

The current efforts to reduce devices' dimensions towards the nanoscale cannot be fully reached without innovative solutions to their power supply. For many applications such as biomedical, deployable sensor networks, or autonomous nanomachines, replacement of exhausted batteries is not an option and wireless devices are desirable or even required.\cite{bioApp,WLesssensors,revBook,revBistable,revPiezo}
In this context, vibrational energy harvesting is attracting considerable attention as vibrations are pervasively available in different environments.\cite{bioApp,WLesssensors,revBook,revBistable,revPiezo,PRLnonlinear,Triboimpact,TriboBroad,PRLDdots,PRLnoiseRect}
A severe limitation of most of the proposed vibrational energy harvesters is their narrow bandwidth of operation at acceptable performance.
Indeed, this has driven an active area of research in recent years.
\cite{revBistable,PRLnonlinear,Triboimpact,TriboBroad,PRLDdots,PRLnoiseRect}
The problem is rendered even more complicated for true nanoscale energy harvesters, i.e. when the dimensions of the whole device lay in the nanoscale. There, quantum mechanical effects may become important. Moreover, very low output voltages are expected, which would prevent the use of conventional electric rectifiers.

Nanogenerators made of piezoelectric nanorods have been proposed for nanoscale energy harvesting some time ago.~\cite{wang2008,sheng2010} When nanorods are subjected to an external force a deformation occurs and this causes an electrical field inside the structure. On the other hand, under the appropriate conditions, a Schottky contact can be formed between the counter-electrode and the tip of the nanorod. Both effects can be used, through a proper design of the device, to generate direct currents. It has been proven that these devices can successfully produce electric power from different sources of vibrations.~\cite{wang2008} However, even in this case, there is a minimum amplitude of the motion of the nanorods needed to produce an efficient rectification.

In this work, we study a mechanism that can convert kinetic energy into electrical work at the nanoscale, which is potentially useful for vibrational energy harvesting.
The proposed mechanism precludes the use of electric rectifiers of any kind. Moreover, it does not require a tuning of the resonances of the system to the main contributions of the vibrational spectrum of the environment, as is the case for most vibrational energy harvesters. Our proposal is based on the long-time behavior of quantum pumping~\cite{ZPBButtiker,Brouwer,avron2000,watson2003,foa2005,strass2005,splettstoesser2005,arrachea2006,nakajima2016,schweizer2016} induced by damped vibrational modes. The idea is that mechanical excitations, such as sound waves traveling through the system or similar phenomena, triggers the movement of a device that hits a conductor. The kinetic energy of the impact is then transformed directly into an electric current through vibrational-induced quantum pumping. The whole process has a nonvanishing direct current component at long times which depends on the geometry of the trajectories in the phase space of the system's normal modes.

This work is organized as follows. In Sec. \ref{sec_theory} we first discuss in more detail the type of processes treated here and then derive the general theory used to describe them. In Secs. \ref{sec_fg} and \ref{sec_fs} we derive for particular (but quite general) cases explicit expressions for the factors needed to evaluate the total charge pumped per hitting event. In Sec. \ref{sec_coupling} we discuss the effect of coupling the proposed devices to a capacitor and derive some limit expressions for the efficiency and the steady-state voltage of operation. In Sec. \ref{sec_quantum} we analyze a simple example that shows how quantum effects can be used to improve the harvester characteristics. Finally, in Sec. \ref{conclusions} we summarize the main conclusions.

\section{General theory\label{sec_theory}}

Before starting with the theory, let us first clarify the type of processes we are dealing with. Our goal is the same as that of macroscopic vibrational energy harvesting but taken to the nanoscale. One wants to extract useful electrical power from ambient residual energies arising from different mechanical excitations.
Those mechanical excitations can emerge in principle from several spontaneous sources such as those produced by biological activities (e.g. walking) or industrial activities (e.g. vibrations stemming from some machinery), but also from sources purposely generated by an external agent as a way of feeding a nanomachine wirelessly.

\begin{figure}
     \begin{center}
         \includegraphics[width=3.3in, trim=0.0in 0.0in 0.0in 0.0in, clip=true]{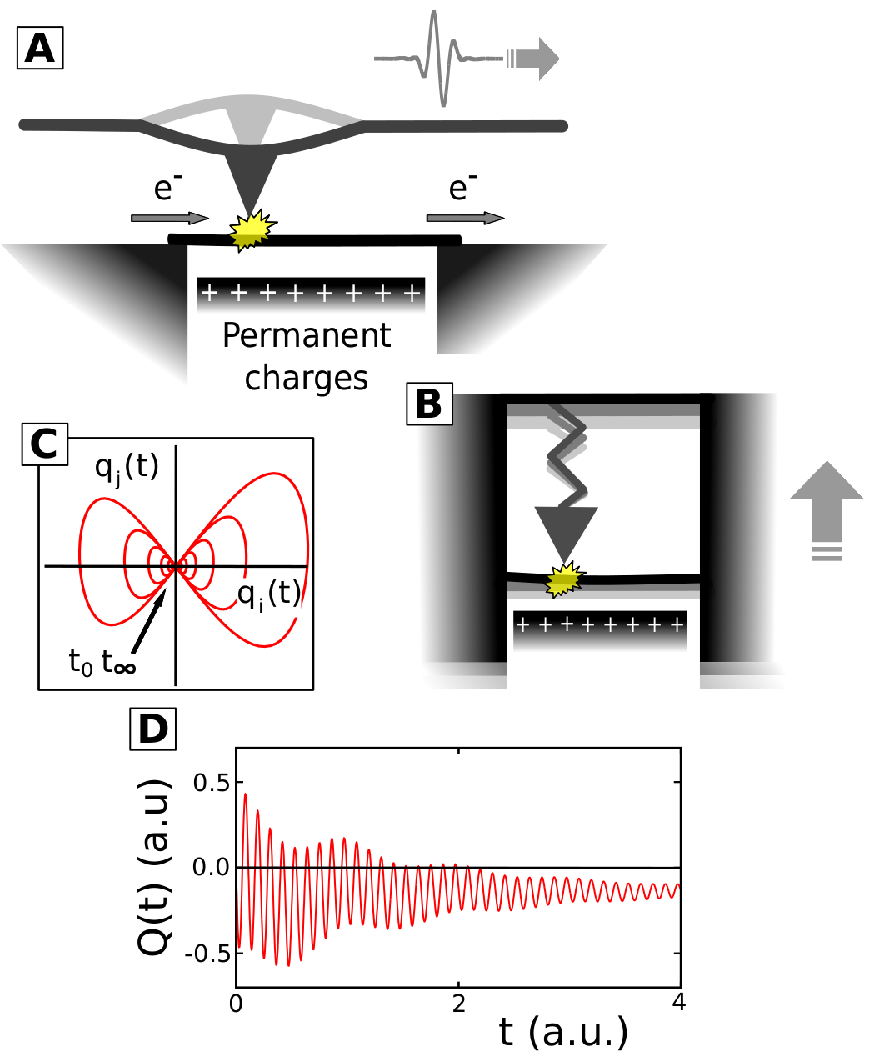}        
    \end{center}
    \caption{\textbf{A} and \textbf{B - } Schemes of the type of systems proposed. An external mechanical force stemming from the environment triggers the movement of a bistable tip (\textbf{A}) or shakes the whole nanodevice (\textbf{B}). As a result of that, a tip hits the system, in this case, a conductive wire suspended over permanent charges. This starts the oscillation of the wire, which in turn pumps electrons between the reservoirs. \textbf{C - } Typical trajectory in the phase space of the normal modes of the wire, represented by $q_i$ and $q_j$. The initial time and the long-time behavior are marked by $t_0$ and $t_\infty$ respectively. The geometry of the trajectories determine the total pumped charge at $t \rightarrow \infty$. \textbf{D - } A typical plot of the pumped charge $Q(t)$ as a function of time $t$, in arbitrary units.}
   \label{fig_scheme}
\end{figure}

The type of systems considered consists of a hitting device that only when, for example, a mechanical wave goes through the device or the whole harvester is shaken, hits in a certain way a conductor connected to two leads.
The motion of the conductor and its coupling to the electronic degrees of freedom is what then pumps current between the reservoirs. This process is depicted in Fig. \ref{fig_scheme}.
We will describe the pumping process quantum mechanically so we are implicitly assuming that the coherence length of the electrons in the conductor is at least of the same order as its characteristic size, which is in the nanoscale.~\footnote{Interesting physical systems where this condition can be found are carbon nanotubes and graphene sheets for example~\cite{foatorres2014}}
In contrast, the motion of the conductor is assumed to be classically treatable.

The starting point of our theoretical description is the well-known formula due to Brouwer, B\"uttiker, Thomas, and Pr\^etre \cite{Brouwer,ZPBButtiker} of the adiabatic charge pumping, which adapted to our problem reads
\begin{equation}
Q_r = e \int_0^{\infty} dt \left( \sum_i \frac{dn_r}{dq_i} \dot{q_i}
   \right) \label{eq_first}
\end{equation}
Here, $e$ is the charge of the electron, and $Q_r$ is what we call the asymptotic pumped charge (APC) from the reservoir $r$, where ``asymptotic'' refers to the long-time limit of the pumped charge $Q(t)$, i.e. $\lim_{t \rightarrow \infty} Q(t)$. This differs from the usual definition for $Q_r$, referring to the charge pumped per cycle~\cite{Brouwer}. In our case there is not a cycle but a hitting event which is unique in principle.
The modes of the mechanical part of the system are labeled $q_i$, and $\frac{dn_r}{dq_i}$ is the emissivity, defined in the low-temperature limit as
~\footnote{The low-temperature limit of the emissivity is used just for simplicity. For finite temperatures an extra integral should be added to the formulas of the scattering factor as now
\begin{equation}
\frac{dn_r}{dq_i} = - \int \frac{df}{d\varepsilon}{\sum_{\beta, \alpha \in r} \frac{1}{2 \pi} \mathrm{Im} \left[
 \frac{\partial S_{\alpha \beta}}{\partial q_i}  S^{\ast}_{\alpha \beta}     \right]} d\varepsilon,
\end{equation}
where $f$ is the Fermi function
}
\begin{equation}
\frac{dn_r}{dq_i} = \sum_{\beta, \alpha \in r} \frac{1}{2 \pi} \mathrm{Im} \left[
 \frac{\partial S_{\alpha \beta}}{\partial q_i}  S^{\ast}_{\alpha \beta}     \right],
\end{equation}
where $S_{\alpha \beta}$ is the element of the scattering matrix $\boldsymbol S$ that connects a conduction channel $\beta$ belonging to some reservoir, to a conduction channel $\alpha$ belonging to the reservoir $r$ ($S_{\alpha \beta}$ is a transmission amplitude for $\alpha$ and $\beta$ belonging to different reservoirs or a reflection amplitude otherwise).
To obtain a simple expression, we expand the emissivity up to linear order in $q_i$,
\begin{equation}
  Q_r \approx e 
  \sum_i \left. \frac{dn_r}{dq_i} \right|_{q_0}   \int_0^{\infty} \dot{q_i} dt 
  + \sum_{i, j} \left. \frac{\partial}{\partial q_j} \frac{dn_r}{dq_i} \right|_{q_0}  \int_0^{\infty} q_j \dot{q_i} dt \label{eq_Qexpan}.
\end{equation}
We assume the system is initially at rest and all excitations decay at long times to the initial condition, i.e. $q_i(0)=q_i(\infty)$. Then, we can use integration by parts, which gives 
\begin{equation}
\int_0^{\infty}q_i\dot{q_j}dt = -\int_0^{\infty} q_j\dot{q_i}dt ,
\end{equation}
to obtain
\begin{eqnarray}
  Q_r & \approx & \sum_{i < j} 
    \underbrace{
  \sum_{\beta, \alpha \in r} 
  \frac{e}{\pi} \mathrm{Im} \left[  \frac{\partial S_{\alpha \beta}}{\partial q_j}  \frac{\partial S^{\ast}_{\alpha \beta}}{\partial q_i}  \right]_{q_0}
  }_{f_s}
    \underbrace{
  \int_0^{\infty} q_i \dot{q_j} dt. \label{eq_Qtotal}
  }_{f_g}
  \end{eqnarray}
This equation is a generalization of Brouwer's formula\cite{Brouwer} for the multiparametric adiabatic charge pumping. It consists of the summation of the contributions from every pair of normal modes $i$ and $j$ to the pumped charge.
Each contribution is the multiplication of two factors, a scattering factor ($f_s$) and a geometric factor ($f_g$). We will see that $f_g$ is independent of the speed at which trajectories are traversed and only depends on their geometry, hence the name. 
Its dependence on the geometry can be written as an enclosed area, but this ``enclosed area'' is something more complex than that of conventional adiabatic quantum pumping~\cite{Brouwer}. As can be seen in Fig. \ref{fig_scheme}-C, for each pair of parameters $q_i$ and $q_j$, there are infinite enclosed areas whose signs depend on the direction in which the trajectories are being traveled. The sum of all these areas gives the geometric factor for the pair ($q_i$,$q_j$) which is not zero in general.~\footnote{To see that, take the integral $\int_0^{\infty} q_i \dot{q}_j dt$ and divide it into time intervals that correspond to the different closed trajectories,
$\int_0^{\infty} dt = \int_0^{t_1} dt+...+\int_{t_i}^{t_{i+1}} dt+...$.
Then simply change the variables of the integrals as $\int_{t_i}^{t_{i+1}} q_i\dot{q}_j d t = \oint q_i(q_j) d q_j$. The last integral is the area enclosed by the particular segment $(i,i+1)$ of the total trajectory
} It is fair to note here that, as well as for conventional quantum pumping, the APC is a first order effect in an expansion of the emissivity as can be noticed in Eq. \ref{eq_Qexpan}. Thus, large pumped currents should not be expected in general.

An interesting aspect of the kind of pumping treated in this work is that it does not need
an external agent that continuously moves the parameters in a certain way. Instead, asymptotic quantum pumping only requires an appropriate initial condition and a damping mechanism, which should always be present in any system. One advantage of dispensing with continuous electrical driving and relying instead on a mechanical triggering is that the displacement currents that make experiments with quantum pumping so difficult are absent here.
\footnote{\label{note1}Displacement currents arise from the capacitive coupling of time-dependent gate voltages with the reservoirs. These currents typically hinder the detection of pumping currents.\cite{brouwer2001}. In our case, gate voltages are not necessary in general but even in the case of using them, as may be the case for proposals similar to those shown in Fig. \ref{fig_scheme}, they are time independent. The time dependence is in the deformation of the system itself, which is independent of any external agent.}
This may open the door to a new way of experimentally studying quantum pumping.

\section{Geometric factor for damped harmonic systems\label{sec_fg}}

\subsection{Impulsive initial conditions}

Let us assume the classical $q_i$ modes correspond to the normal modes of a system initially at rest that suffered an impulsive initial condition.
For the moment, let us also assume that temperature is zero.
Then, we can write
\begin{equation}
q_i (t) = a_i \sin (\omega_i \omega_0 t) e^{-\gamma_i \omega_0 t} \label{eq_q}
\end{equation}
where $t$ is the time, $\omega_i$ is the resonant frequency of the normal mode $i$ in units of a reference frequency $\omega_0$, and $\gamma_i$ is the damping factor, also in units of $\omega_0$.
Note that this equation makes explicit the meaning of the long-time limit of $Q_r$, $t \gg \max{[1/(\gamma_i \omega_0)]}$.
The value of the $a_i$ coefficients depend on the initial velocities of each normal mode, $a_i=\dot{q}_i (0)/(\omega_i \omega_0)$, which, in turn, depends on the details of how the tip hits the system.
Integrating the geometric factor $f_g$ with $q_i$ given by Eq. \ref{eq_q} yields
\begin{equation}
f_g(i,j) = \frac{ (a_i
  a_j) \omega_i \omega_j  [ (\omega_i^2 - \omega_j^2) + (\gamma_i^2 -
  \gamma_j^2)]}{[ (\gamma_i + \gamma_j)^2 + (\omega_i^2 + \omega_j^2)]^2 - 4
  \omega_i^2 \omega_j^2} \label{eq_fg}
\end{equation}
Note that $f_g$ is independent of $\omega_0$, which gives the time scale of the whole process. 
Thus, $f_g$ only depends on the geometry of the trajectories, given by the pairs $(a_i,a_j)$, $(\gamma_i,\gamma_j)$, and $(\omega_i,\omega_j)$.
From Eq. \ref{eq_fg}, it is clear that completely random initial conditions, which would correspond to random values of $a_i$ and $a_j$, would make the average value of $f_g$ zero.
This highlight the obvious fact that it is not possible to extract energy from thermal fluctuations (if the whole system is described by a unique temperature).
However, if the tip is moved by an external source, see the discussion at the beginning of sec. \ref{sec_theory}, its shape is kept constant between hitting events, and it hits the device at the same position, all the ratios $a_i/a_j$ will be the same and only the absolute values of the $a_i$ coefficients will change.
If this is the case, then, the APC can only change its magnitude but not its sign between hitting events.
Fig. \ref{fig_scheme} shows schemes of two possible setups of the system.
There, a tip hits a conducting wire randomly in time but always at the same place and from the same direction. Different shapes of the tip or multiple tips can also be used to control which normal modes of the wire will be excited.

In Eq. \ref{eq_fg} one can check that decreasing the damping factors increases the total pumped charge. However, there is an upper limit to the APC, given by
\begin{equation}
  \lim_{\gamma \rightarrow 0}  \int_0^{\infty} q_i \dot{q_j} dt = \pm
  \frac{(a_i a_j)}{2}  \sqrt{\left( \frac{1 + \left( \frac{\omega_j}{\omega_i}
  \right)^2}{1 - \left( \frac{\omega_j}{\omega_i} \right)^2} \right)^2 - 1}
\end{equation}
where $\pm$ corresponds to $\omega_i \gtrless \omega_j$.
\begin{figure}
     \begin{center}
         \includegraphics[width=3.3 in, trim=0.0in 0.0in 0.0in 0.0in, clip=true]{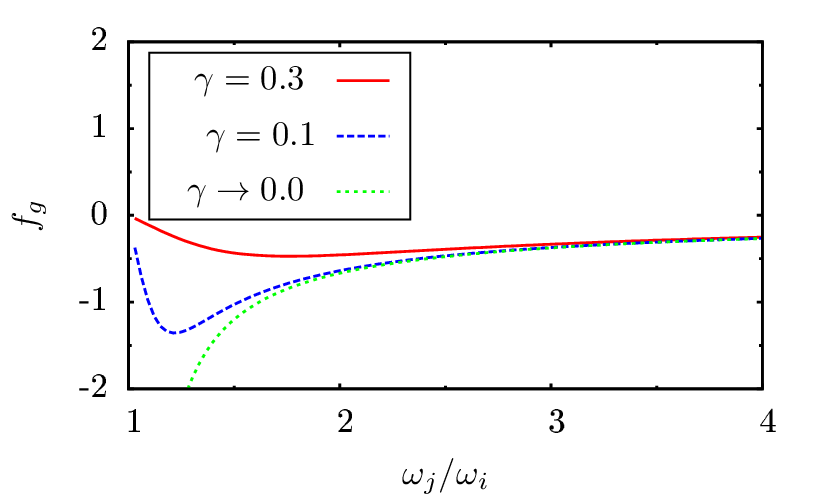}        
    \end{center}
    \caption{Geometric factor $f_g$, in units of $(a_i a_j)$, as a function of the ratio between the frequencies of two normal modes, $\omega_j/\omega_i$ and for different damping factors, $\gamma_i=\gamma_j=\gamma$.}
   \label{fig_geometric-factor}
\end{figure}
Fig. \ref{fig_geometric-factor} shows the dependence of $f_g$ on the damping factors and the frequency ratios between modes. We can see that the closer the frequencies of two modes, the larger their contribution to the pumped charge. Then, considering two consecutive modes, which will give the largest contribution, the higher their frequency the better.

Up to this point we have only considered the zero temperature case for the geometric factor, which implies $q_i(t=0)=0$ and the absence of stochastic forces in the trajectories. To address the effect of the temperature we will consider a more realistic situation where the dynamics of $q_i(t)$ is determined by a Langevin-like equation
\begin{equation}
\ddot{q}_i= -\left ( \omega_i^2 + \gamma_i^2 \right ) q_i - 2 \gamma_i \dot{q}_i + \xi_i . \label{eq_qLangevin}
\end{equation}
Here, $\xi_i$ accounts for the stochastic forces . These forces have zero mean $\left < \xi_i \right > = 0$ and are assumed local in time with a correlation function given by $\left < \xi_i(t) \xi_i(t') \right > = D_i \delta(t-t')$, where $D_i$ is chosen such as to fulfill the fluctuation-dissipation theorem, $D_i=2 K T \gamma_i$. At zero temperature and for impulsive initial conditions one recovers Eq. \ref{eq_q}. 
We are assuming that the hitting device is an object large enough so that its dynamics is not affected by thermal noise. Therefore, only when, for example, some mechanical wave goes through the system or the whole harvester is shaken, the hitting device is triggered. 
We also assume that the impact is fast compared with the time scales of the vibrational modes coupled to the electronic degrees of freedom. 
Then, the only role of the hitting device is to provide the impulsive initial condition. For that reason, its dynamics will not be considered explicitly.
\begin{figure}
     \begin{center}
         \includegraphics[width=3.3in, trim=0.0in 0.0in 0.0in 0.0in, clip=true]{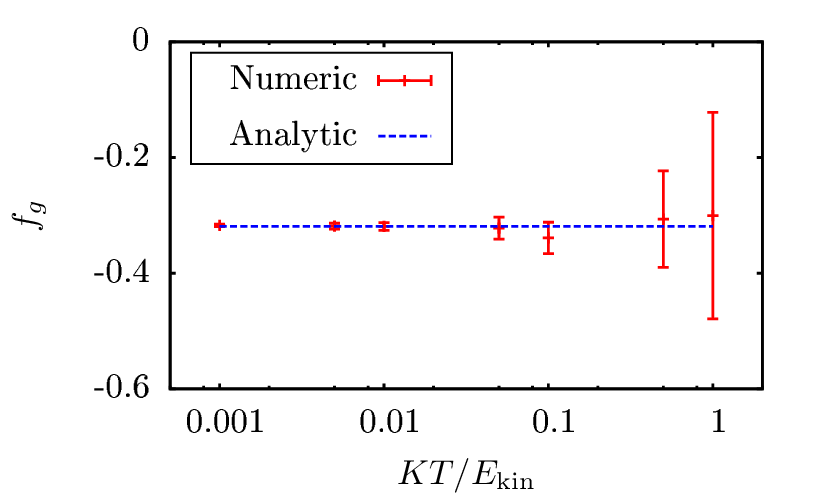}        
    \end{center}
    \caption{Effect of the temperature on $f_g$. $KT$ is the Boltzmann constant times the temperature. $E_{\mathrm{kin}}$ is the kinetic energy added by the impulsive initial condition. The line marked as ``analytic'' corresponds to Eq. \ref{eq_fg} ($KT=0$). The error bars shown as ``numeric'' are centered at the average value of $f_g$ obtained numerically for finite temperatures. The width of the error bars corresponds to $2 \sigma/\sqrt{N}$ where $\sigma$ is the standard deviation of the set of trajectories at the same temperature and $N$ is the number of trajectories run ($N=100$). $f_g$ is in units of $(a_i a_j)$. See text for details.
    }
   \label{fig_efectKT}
\end{figure}

We numerically solved Eq. \ref{eq_qLangevin} for two modes with $\omega_2 = 2 \omega_1$ and $\gamma_1=\gamma_2=0.1 \omega_1$. The geometric factor, Eq. \ref{eq_Qtotal}, was numerically evaluated 
using a final time equal to $10/\gamma_1$. The initial position and velocity of the modes were chosen from a thermal ensemble and then at $t=0$ a quantity equal to $\sqrt{2 E_{\mathrm{kin}}}$ was added to the initial velocities, where $E_{\mathrm{kin}}=\dot{q}_i^2/2=0.5$ in arbitrary units ($q_i$ is in units of $a_1$ and $\dot{q_i}$ is in units of $a_1 \omega_1$). 
Fig. \ref{fig_efectKT} shows the average value of the geometric factor (and its error) obtained from the simulations as a function of the temperature. As can be seen, the only role of temperature is to broaden the distribution functions of $f_g$ around the values predicted by Eq. \ref{eq_fg}.

\subsection{Displacive initial conditions\label{sec_fg_dis}}
\begin{figure}
     \begin{center}
         \includegraphics[width=2.8in, trim=0.0in 0.0in 0.0in 0.0in, clip=true]{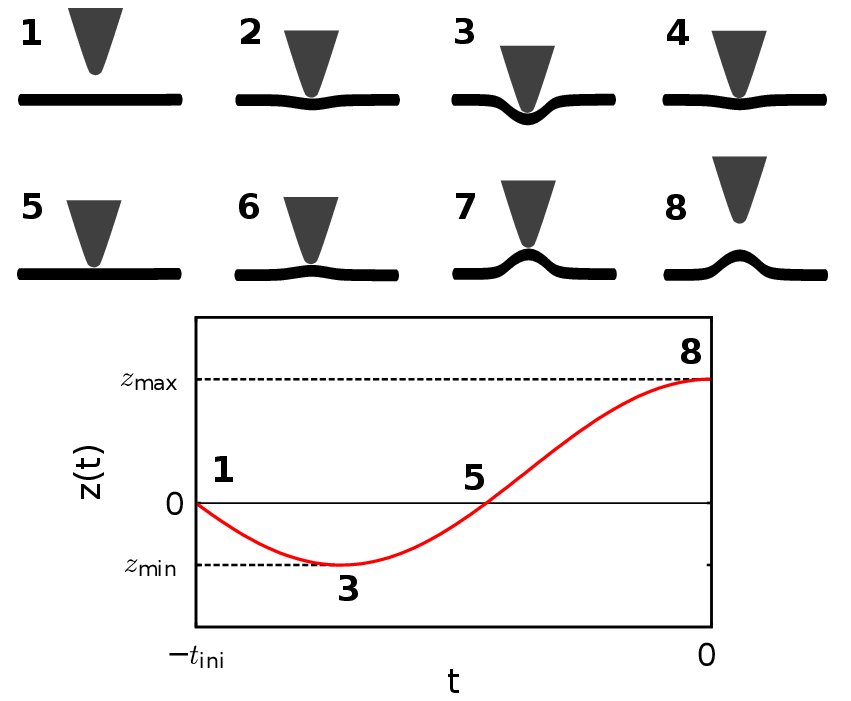}        
    \end{center}
    \caption{Scheme of the type of processes that can give rise to displacive initial conditions. Due to some external excitation a tip first pushes a conductive wire and then it retreats but pulling the wire with it in the process as consequence of van der Waals forces. The movement of the conductive wire during the process depicted and its subsequent free movement, after the interaction finished, pump current between two reservoirs. In the plot $z(t)$ represents the position of the tip with respect to the wire.}
   \label{fig_scheme-dis}
\end{figure}

\begin{figure}
     \begin{center}
         \includegraphics[width=3.3in, trim=0.0in 0.0in 0.0in 0.0in, clip=true]{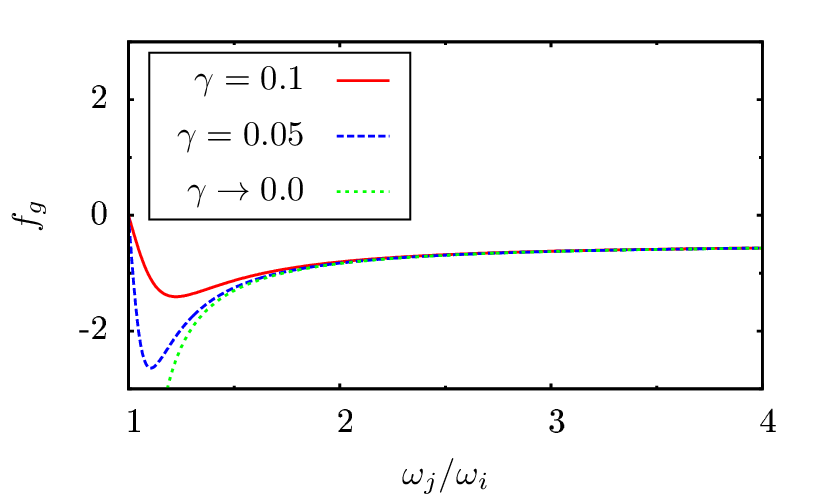}        
    \end{center}
    \caption{Same as Fig. \ref{fig_geometric-factor} but for displacive initial conditions. $f_g$ is in units of $(q_{0 i} q_{0 j})$.}
   \label{fig_gf_displasive}
\end{figure}

In this subsection we analyze a complementary case to that studied in the previous subsection. 
In the displacive initial conditions, the velocities of all normal modes are zero at the beginning of the free movement but not the positions.
A physical situation corresponding to this case may be, for example, a tip that first pushes a conductor and then, when moving back, pulls the conductor with it, due to the van der Waals forces.
At some point, the restoring forces overcome the van der Waals forces and the conductor is released, marking the beginning of its free motion.
This situation is depicted in Fig. \ref{fig_scheme-dis}.
Note that here the ``collisional'' time can be large compared with the system's dynamics.
The equation of motion for the normal mode $q_i$ can be written in this case as
\begin{equation}
 q_i (t) =  \left \{
 \begin{array}{ll}
  a_i z(t)  & ~ ~ \mathrm{for~} -t_{\mathrm{ini}} < t < 0 \\
  q_{0i} e^{- i \gamma_i \omega_0 t} \cos \left ( \omega_i \omega_0 t \right ) 
 & ~ ~ \mathrm{for~} 0 \leq t < \infty
 \end{array}
\right . \label{eq_q_dis} ,
\end{equation}
where $z(t)$ is a coordinate that describes the tip's movement, $a_i$ is the weight of the $z$ coordinate on the normal mode $q_i$, and $-t_{\mathrm{ini}}$ marks the beginning of the interaction between the tip and the conductor. We will describe the tip's movement by the minimal expression
\begin{equation}
 z(t) = \frac{( z_{\mathrm{max}} - z_{\mathrm{min}})}{2} \cos\left( \omega_z t \right)
 + \frac{(z_{\mathrm{max}} + z_{\mathrm{min}})}{2} \label{eq_zt} ,
\end{equation}
where $z_{\mathrm{max}}$ and $z_{\mathrm{min}}$ are the maximum and minimum values of $z(t)$ respectively (while the tip is still in contact with the conductor). From Eq. \ref{eq_zt} it is clear that $z_{\mathrm{max}}=z(0)$ and $q_{0i}=a_i z_{\mathrm{max}}$ The value of $\omega_z$ is calculated so that $z(-t_{\mathrm{ini}})=0$, i.e. 
\begin{equation}
\omega_z = \left ( 2 \pi - \arccos \left [
\frac{(z_{\mathrm{max}} + z_{\mathrm{min}})}
{(z_{\mathrm{min}} - z_{\mathrm{max}})} \right ] \right ) /t_{\mathrm{ini}} 
\end{equation}
The integration of the geometric factor is now split into two parts
\begin{equation}
 f_g = \int_{-t_{\mathrm{ini}}}^\infty q_i \dot{q}_j d t
 = \int_{-t_{\mathrm{ini}}}^0 q_i \dot{q}_j d t + \int_0^\infty q_i \dot{q}_j d t
\end{equation}
The first integral is easy to evaluate, it gives $q_{0 i} q_{0 j}/2$, while the second one is more cumbersome. The final result is
\begin{widetext}
\begin{equation}
f_{g} = q_{0 i} q_{0 j} \left[\frac{1}{2}-\frac{\gamma_{j}\left(\gamma_{i}+\gamma_{j}\right)\left[\left(\gamma_{i}+\gamma_{j}\right)^{2}+\omega_{i}^{2}\right]+\left[\left(\gamma_{i}+\gamma_{j}\right)\left(\gamma_{i}+2\gamma_{j}\right)-\omega_{i}^{2}\right]\omega_{j}^{2}+\omega_{j}^{4}}{\left[\left(\gamma_{i}+\gamma_{j}\right)^{2}+\omega_{i}^{2}\right]^{2}+2\left(\gamma_{i}+\gamma_{j}-\omega_{i}\right)\left(\gamma_{i}+\gamma_{j}+\omega_{i}\right)\omega_{j}^{2}+\omega_{j}^{4}}\right]
\label{eq_fg_dis}
\end{equation}
\end{widetext}
The dependence of $f_g$ with $\gamma$ and the ratio $\omega_i/\omega_j$  is similar to that of impulsive initial conditions. See Fig. \ref{fig_gf_displasive}. The limit of small $\gamma$ is now
\begin{equation}
\lim_{\gamma\rightarrow\infty}f_{g} = q_{0 i} q_{0 j} \frac{1}{2}\frac{\left[1-\left(\frac{\omega_{j}}{\omega_{i}}\right)^{4}\right]}{\left[1-\left(\frac{\omega_{j}}{\omega_{i}}\right)^{2}\right]^{2}}
\end{equation}

\begin{figure}
     \begin{center}
         \includegraphics[width=3.3in, trim=0.0in 0.0in 0.0in 0.0in, clip=true]{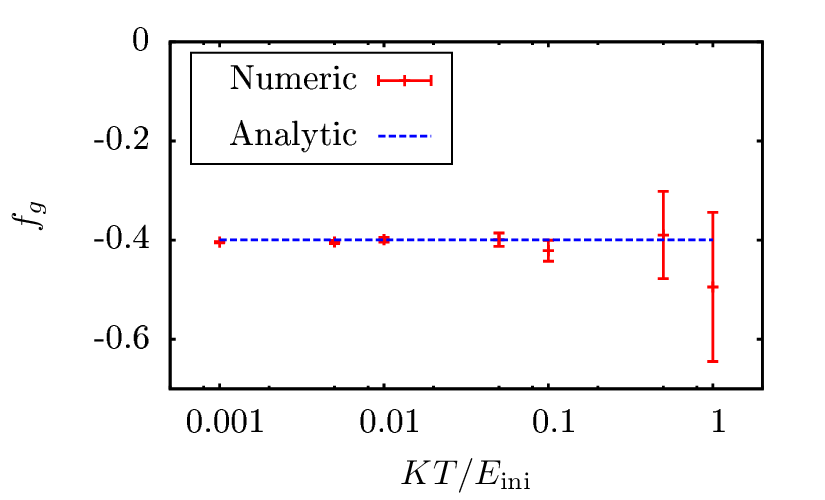}        
    \end{center}
    \caption{Same as Fig. \ref{fig_efectKT} but for displacive initial conditions. $KT$ is in units of the initial potential energy $E_{\mathrm{ini}}=(\omega_i^2+\gamma^2) q_{0 i}^2/2$ and $f_g$ is in units of $(q_{0 i} q_{0 j})$.}
   \label{fig_efectKT_dis}
\end{figure}

As in Subsec. \ref{sec_fg_dis} we implicitly assumed $KT=0$ in Eqs. \ref{eq_q_dis} and \ref{eq_fg_dis}. However, the effect of the temperature is the same as before. It only broadens the distribution function of $f_g$ around the value predicted by the zero-temperature formulas. This can be seen in Fig. \ref{fig_efectKT_dis} where we performed the same type of calculation as that described in the previous subsection, see the text around Eq. \ref{eq_qLangevin}.

\section{Scattering factor of an oscillating wire\label{sec_fs}}

To analyze the effect of the scattering factor $f_s$  we need to resort to particular examples. Let us examine the case of a conductive wire suspended over an electret material
\footnote{An electret is a dielectric material that has a quasi-permanent electric charge or dipolar polarization. An example of its application for energy harvesting can be found in Ref. ~\onlinecite{electret}.}
as shown in panels A and B of Fig. \ref{fig_scheme}.
For simplicity, we assume a small capacitive coupling between the electrons of the wire and the permanent charges.
The potential $U$ sensed by the electrons traversing the wire can now be taken as $U(x)= U_0 z(x)$, where $z$, the separation between the wire and the electret material, depends on the position $x$ along the wire.
Then, the electronic Hamiltonian, written in term of the transverse normal modes of the wire, reads
\begin{equation}
  \hat{H} = \frac{\hat{p}^2}{2 m_e} + U_0 \sum_i q_i ( t) \sin \left( \frac{2 \pi \omega_i  x}{L} \right)
  \Theta(x) \Theta(L-x), \label{eq_H}
\end{equation}
where $\hat{p}$ and $m_e$ are the momentum and mass of the electron, $\Theta$ is the Heaviside step function, $L$ is the length of the wire, $q_i$ is the amplitude of the $i$-th normal mode of the wire (considered in this approximation as a classical variables), and $\omega_i$ is in this case an integer between $1$ and $\infty$.
Note that, for simplicity, we excluded the electron's spin of the analysis.
To solve our problem, we start by first noticing that our Hamiltonian is of the form
$\hat H =  \hat{p}^2/(2 m_e) + \sum_i U_{q_i}$.
If we define $S$ and $S_{q_i}$ as the scattering matrices associated with the Hamiltonians $\hat H$ and ${\hat H}_{q_i}$ respectively, where $\hat H_{q_i} =  \hat{p}^2/(2 m_e) + U_{q_i}$, then, by using the Fisher and Lee formula,~\cite{Fisher1981} one finds
\begin{equation}
\left . \frac{\partial \boldsymbol{S}^{}} {\partial q_i} \right |_{q_0} =
\left . \frac{\partial \boldsymbol{S}_{q_i}^{}} {\partial q_i} \right |_{q_0} \label{eq_dSq=dS}
\end{equation}
where $q_0=q_j(t_0)=0$. See App.~\ref{app_A}.
${\hat H}_{q_i}$ is the same Hamiltonian than that presented in Refs.~\onlinecite{PRLBustos,PRBLucas} for the
Thouless motor. As shown there, one can obtain analytically the scattering matrix of the problem by linearizing the Hamiltonian for momenta close to $\hbar k_i=\pm \hbar \pi \omega_i/L$. See App. \ref{app_B}. Using this result, we obtain the derivative of the scattering matrix of the original problem, Eq. \ref{eq_H},
\begin{eqnarray}
\left . \frac{\partial \boldsymbol{S}^{}} {\partial q_i} \right |_{q_0} =- \frac{U_0 L}{2 \hbar v_F} \mathrm{sinc} \left ( \Delta E_i \right ) e^{i \Delta E_i } \boldsymbol{\sigma}_z \label{eq_dS}
\end{eqnarray}
where $\Delta E_i= \left ( \frac{L}{\hbar v_F} \right ) \left (\varepsilon-\frac{\hbar^2 k_i^2}{2 m_e}\right )$, $\boldsymbol{\sigma}_z$ is the ``$z$'' Pauli matrix, $v_F$ is the Fermi velocity, and $\varepsilon$ is the Fermi energy.
The scattering factor $f_s$ is obtained by assuming the momentum of the electron is close to both $\hbar k_i=\pm \hbar \pi \omega_i/L$ and $\hbar k_j=\pm \hbar \pi \omega_j/L$. Then, one can apply Eq. \ref{eq_dS} to the derivatives with respect to $q_i$ and $q_j$. This results in
\begin{eqnarray}
f_s(i,j,\varepsilon) & = & \frac{e}{\pi} \left (    \frac{U_0 L}{2 \hbar v_F} \right )^2
   \mathrm{sinc} \left ( \Delta E_i \right )
   \mathrm{sinc} \left ( \Delta E_j \right ) \notag  \\
   &&
   \times \sin \left ( \Delta E_i-\Delta E_j \right ) . \label{eq_fs}
\end{eqnarray}
As can be noticed, the scattering factor $f_s(i,j,\varepsilon)$ depends on the Fermi energy and the pair of modes $i$ and $j$ under consideration. Taking its maximum value for each pair of $(i,j)$ modes, one can check that pairs of modes with the closest frequencies, $\omega_j=\omega_i+1$ for $j>i$, give the maximum contribution to the APC, Eq. \ref{eq_Qtotal}. One can also check that, among those pairs with $\omega_j=\omega_i+1$, the ones with the lowest frequencies, smallest $\omega_i$, give the largest contribution to APC. This is the opposite of the behavior of $f_g$ discussed in the previous section.

\begin{figure}
     \begin{center}
         \includegraphics[width=3.3 in, trim=0.0in 0.0in 0.0in 0.0in, clip=true]{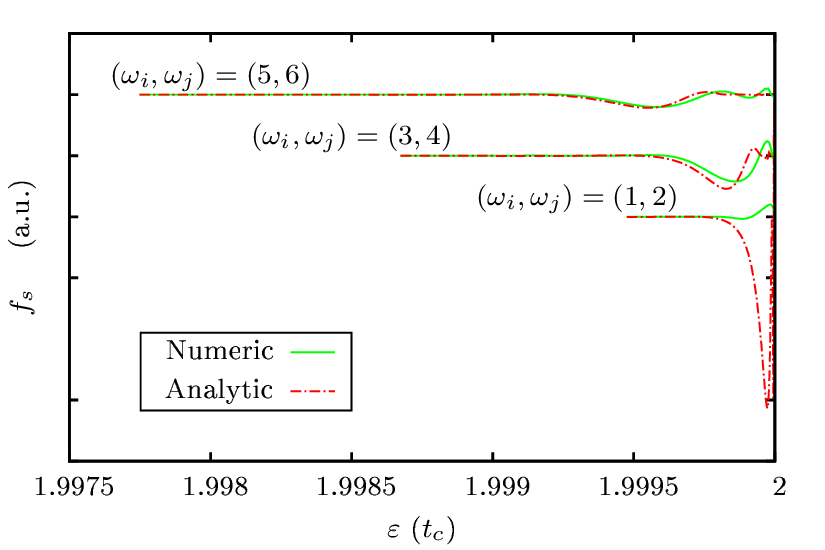}        
    \end{center}
    \caption{Comparison of the scattering factor $f_s$, in arbitrary units, evaluated using Eq. \ref{eq_fs} and numerically. $\varepsilon$ is the Fermi energy in units of $t_c$, the coupling constant of the tight-binding chain. See text for details.}
   \label{fig_comparison}
\end{figure}

The above result was confirmed by numerical calculations based on a tight-binding model. Important deviation were observed only for the smallest ($\omega_i$)s, where Eq. \ref{eq_fs} overestimate the maximum value of $f_s(\varepsilon)$, see Fig. \ref{fig_comparison}.
The tight-binding model\cite{PRBLucas,RevMex} used in the figure consisted of a linear chain of 400 sites with site energy $E_n= U_0 \sum_i q_i ( t) \sin \left( \frac{2 \pi \omega_i  n}{L} \right )$, where $L=400$, $U_0=0.1$, and $q_i$ is given by Eq. \ref{eq_q}. Only first neighbors couplings were considered with a coupling constant $t_c=1$, thus setting the energy scale. Leads were attached to sites $n=1$  and $n=400$ with a coupling constant equal to $t_c$. The self-energies of the leads were taken as
\begin{equation}
\Sigma (\varepsilon ) = \lim_{\eta \rightarrow 0^+} \frac{\varepsilon+\mathrm{i}\eta }{2}
-\mathrm{sgn}(\varepsilon)\sqrt{\left( \frac{\varepsilon+\mathrm{i}\eta }{2}\right)
^{2}-t_c^{2}} , 
\end{equation}
where $\varepsilon$ is the Fermi energy. The numerical value of $f_s$ was obtained from the numerical derivative of the scattering matrix around $q_0$. The scattering matrices were calculated from the retarded Green's functions as shown in App. \ref{app_A} and Refs.~ \onlinecite{Fisher1981,RevMex,cattena2014,PRBLucas}.

To compare the maximum contribution that each pair of modes may have to the APC we rewrite Eq. \ref{eq_Qtotal} as
\begin{equation}
Q(\varepsilon) = \left (\frac{e}{\pi} \frac{U_0^2 L^2}{4 \hbar^2} \right ) \sum_{i<j} \left ( a_i a_j \right ) N_{i,j}(\varepsilon) , \label{eq_Nij}
\end{equation}
where $N_{i,j}(\varepsilon)= f_s(i,j,\varepsilon) f_g(i,j)$ with 
$f_g(i,j)$ and $f_s(i,j,\varepsilon)$ given by Eqs. \ref{eq_fg} and \ref{eq_fs} respectively, but
without the prefactors $\left (\frac{e}{\pi} \frac{U_0^2 L^2}{4 \hbar^2} \right )$ and $\left ( a_i a_j \right )$.
Note we are assuming impulsive initial conditions for the calculation of $f_g$.
The quantity $N_{i,j}(\varepsilon)$ is independent of the initial conditions, given by the value of the pairs $(a_i,a_j)$, but it still depends on $\varepsilon$.
Thus, we define $N^{max}_{i,j}$ as the maximum value of $N_{i,j}(\varepsilon)$ allowed by a variation of $\varepsilon$.

\begin{figure}
     \begin{center}
         \includegraphics[width=3.3 in, trim=0.0in 0.0in 0.0in 0.0in, clip=true]{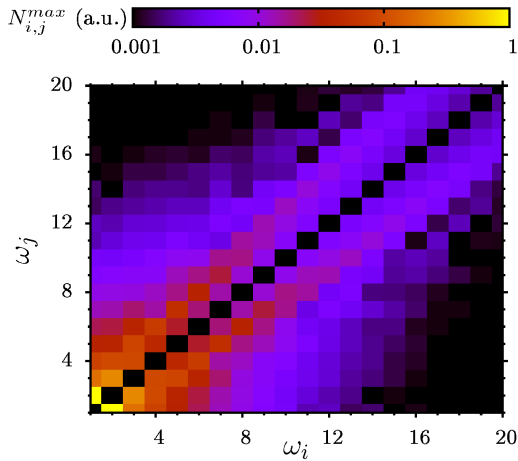}        
    \end{center}
\caption{Maximum relative contribution that each pairs of modes $\omega_i$ and $\omega_j$ may have to the APC. $N^{max}_{i,j}$ is the maximum value of $N_{i,j}(\varepsilon)$, see Eq. \ref{eq_Nij}, obtained by varying $\varepsilon$. Impulsive initial conditions were assumed for the geometric factor in the limit $\gamma \rightarrow 0$.} 
   \label{fig_Qmax-FP}
\end{figure}

Fig. \ref{fig_Qmax-FP} shows the value $N^{max}_{i,j}$, in arbitrary units, for different pairs of modes $i$ and $j$. For simplicity in the figure we assumed $v_F=\sqrt{2 \varepsilon /m_e}$. We used $L=100nm$, although normalized figures are indistinguishable with respect to a variation of $L$ or $m_e$.
In the figure, we can see that the lowest frequency modes give the largest contribution to $Q$. This may have important consequences for the design of the proposed device as it can help to optimize the  hitting mechanism.

\section{Coupling to a capacitor\label{sec_coupling}}

We have shown that it is possible to harvest mechanical energy from the environment by using geometric rectification. However, this energy has to be stored into a voltage bias, and now the problem is to understand the back action of it on the pumping process.
Let us assume our system is connected in series with a capacitor with capacitance $C$ and let us 
simplify the analysis by considering only small voltages.
Then, the total charge accumulated in the capacitor $Q_{_R}^{\mathrm{total}}$ produces a voltage bias $V$ according to $ V = Q_{_R}^{\mathrm{total}} / C$, where $V=V_L-V_R$ with $L$ and $R$ labeling the left and right leads respectively.
The voltage bias induces, in turn, an additional force $F_i$, given by\cite{PRLBustos,PRBLucas}
\begin{eqnarray}
F_i = \left( \frac{d n_{_L}}{d q_i} - \frac{d n_{_R}}{d q_i}
   \right) \frac{e V}{2},
\end{eqnarray}
and this force will affect the dynamics of the entire system.
In principle, this force could change the equilibrium positions and the normal modes of the system between hitting events or, even worse, while the system is relaxing. This is because $V$ varies with time.
The variation of $V$ with time is a consequence of the charge accumulation driven by charge pumping, Eq. \ref{eq_Qtotal}, and the charge leakage due to the bias current $I^{\mathrm{bias}}$. The bias current can be described by $I_L^{\mathrm{bias}}=\frac{2 e^2}{h}T_{_{L R}} V$ where $T_{_{L R}}$ is the transmittance, and the factor $2$ takes into account the spin multiplicity. 
A full treatment of the problem then requires the solution of an additional coupled equation,
\begin{eqnarray}
 V ( t) & = & - \int_0^t \frac{2 e^2 T_{_{L R}} V}{h C} d t' - \notag \\
   & & \int_0^t \sum_i \frac{1}{2 C} \left( \frac{d n_{_L}}{d q_i} -
   \frac{d n_{_R}}{d q_i} \right)  \dot{q}_i d t', \label{eq_Vcomplex}
\end{eqnarray}
where the time $t$ can be large enough as to include several ``hitting'' events of the type described by Eq. \ref{eq_Qtotal}. 
To gain some understanding of the role of current-induced forces without resorting to numerical simulations, we will make additional assumptions. First, the hitting events are random but sufficiently far apart such that they do not interfere with each other. Second, after waiting enough time such that a large number of hitting events have occurred, a steady state is reached where the variation of $V(t)$ is small compared with its mean value $\left < V \right >$. The latter is a good approximation when the average pumped charge during a hitting event and the total charge leaked between events are both negligible compared with the total charge accumulated in the capacitor. This condition can be written as $\left ( 2 e^2 T_{_{L R}}\right) / \left (h \nu C \right ) \ll 1$ where $\nu$ is the frequency of events that lead to APC.
Considering the above, we can clear the mean voltage from Eq. \ref{eq_Vcomplex} giving,
\begin{eqnarray}
\left < V \right > & \approx & \frac{h \langle Q_{_R} \rangle \nu }{2 e^2 T_{_{L R}}} \label{eq_Vsimple}
\end{eqnarray}
where $\langle Q_{_R} \rangle$ is the mean value of APC. Using Eq. \ref{eq_Vsimple} and expanding the emissivity up to linear order in $q_i$ (similarly to what we did in Eq. \ref{eq_Qexpan}), we obtain a simple expression for the current-induced forces,
\begin{eqnarray}
 F_i & = & \left( \left. \frac{dn_{_L}}{dq_i} \right|_{q_0} -
   \left. \frac{dn_{_R}}{dq_i} \right|_{q_0} \right) e \frac{\left < V \right >}{2}^{} \notag \\ &&
   + \sum_j
   \left( \left. \frac{\partial}{\partial q_j}  \frac{dn_{_L}}{dq_i}
   \right|_{q_0} - \left. \frac{\partial}{\partial q_j}  \frac{dn_{_R}}{dq_i}
   \right|_{q_0} \right) e \frac{\left < V \right >}{2} q_j \label{eq_Force}
\end{eqnarray}
The first term just redefines the equilibrium position of the ``$q_i$'' modes, while the second one couples linearly the modes among each other and changes their natural frequency of resonance.
However, the whole system is still harmonic.
Therefore, the expression for $f_g$, Eq. \ref{eq_fg}, remains valid even for finite voltages. 

If we consider an steady-state situation such as that described in the context of Eq. \ref{eq_Vsimple}, we can readily obtain the total work done by the current-induced forces after a hitting event.
The result is simply the energy added to the capacitor
\footnote{To recover the expression for $Q_L$ shown in Eq. \ref{eq_Qtotal} we assumed that there is not an accumulation of charges in the system $Q_L=-Q_R$ and that the equilibrium positions do not change ($q_i(t_0) \approx q_i(t_\infty)$). The latter is reasonable for $Q_L$ much smaller than the total charge accumulated in the capacitor, which implies $V(t_0) \approx V(t_\infty)$}
\begin{equation}
W=\int \boldsymbol{F} \cdot \boldsymbol{dq}= \int \sum_i F_i \dot{q}_i dt=-Q_{_R} \left < V \right >.
\end{equation}

\section{Performance and quantum effects\label{sec_quantum}}
\begin{figure}
     \begin{center}
         \includegraphics[width=3.3 in, trim=0.0in 0.0in 0.0in 0.0in, clip=true]{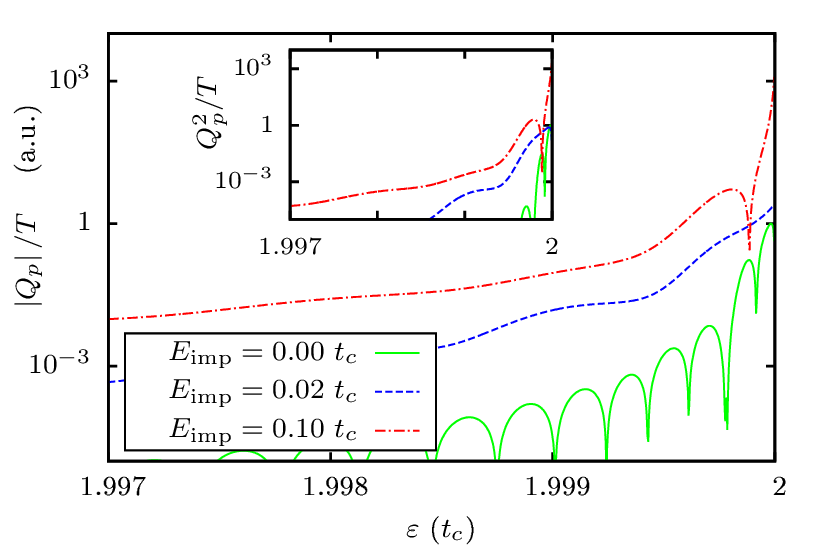}        
    \end{center}
    \caption{Effect of impurities with different energies ($E_{\mathrm{imp}}$) on the ratios $\left | Q_{_R} \right | /T_{_\mathrm{LR}}$ and $Q_{_p}^2/T_{_\mathrm{LR}}$ in arbitrary units. $\varepsilon$ is the Fermi energy.}
   \label{fig_Quan-imp}
\end{figure}
Considering that the energy of the whole process comes from the initial kinetic energy of the hitting device and the fact that we are interested in accumulating energy in a capacitor, it is natural to define the efficiency of the global process as
\begin{eqnarray}
\eta = \frac{\langle Q_{_R} V \rangle}{\langle E_{\mathrm{kin}} \rangle} ,
\end{eqnarray}
where $\langle E_{\mathrm{kin}} \rangle$ is the average initial kinetic energy. Then, assuming the validity of Eq. \ref{eq_Vsimple}, we can write, for impulsive initial conditions,
\begin{eqnarray}
\eta \approx
\frac{h \nu \langle Q_{_R} \rangle^2 }{e^2 T_{_{\mathrm{LR}}}
   \omega^2_0 \sum_i \langle \omega^2_i a^2_i \rangle}. \label{eq_eta}
\end{eqnarray}
Note that the efficiency does depend on the absolute temporal scale of the charge pumping process (proportional to $1/\omega_0$), while neither the APC, Eq. \ref{eq_Qtotal}, nor the steady-state voltage, Eq. \ref{eq_Vsimple}, does.

Several parameters can be tuned to increase $\eta$, but particularly interesting is the ratio $\langle Q_{_R} \rangle/T_{_\mathrm{LR}}$.
In principle, different quantum effects can be used to reduce $T_{_\mathrm{LR}}$. The question is: Will quantum effects also reduce the pumped charge? One of the simplest examples to study this is the use of Anderson's localization induced by impurities in molecular wires.

To study the effect of an impurity on the ratio $Q_{_R}/T_{_\mathrm{LR}}$, we performed a tight-binding calculation of the APC similar to that described in the context of Fig. \ref{fig_comparison}. The defect was placed at site $n=200$ for a chain of 400 sites. The site's energy of the impurity was $E=E_n+E_{\mathrm{imp}}$ and only modes with $\omega_i$ equal to 1 and 2 were excited assuming an impulsive initial condition with $a_1=a_2$. 
The geometric factor was evaluated directly from Eq. \ref{eq_fg} and the rest of the parameters of the tight-binding calculation were the same than those of Fig. \ref{fig_comparison}.
Fig. \ref{fig_Quan-imp} shows the effect of impurities with different energies on the ratios $\left | Q_{_p} \right | /T_{_\mathrm{LR}}$ and $Q_{_p}^2/T_{_\mathrm{LR}}$, the latter shown in the inset. Considering Eqs. 
\ref{eq_Vsimple} and \ref{eq_eta}, the figure shows that quantum-induced localization of the electron's wave function can increase up to three orders of magnitude the energy accumulated in the capacitor and the efficiency of the whole process.
This emphasizes the key role that quantum mechanics may have on nanoscale vibrational energy harvesting.

\section{Conclusions\label{conclusions}}

We have studied a previously unreported mechanism that can turn residual kinetic energy  directly into useful electrical work in the nanoscale by using quantum pumping.
As an application example, we have analyzed a solvable system consisting of a wire suspended over permanent charges where we find the conditions for maximizing the asymptotic pumped charge.
We have discussed the effects of coupling general systems to a capacitor where we include in the analysis the effect of current-induced forces. We have given explicit expressions for the steady-state voltage of operation and the efficiency of the harvesting process in the limit of small but stationary voltages. Finally, we have shown how quantum effects can be used to enhance the performance of energy harvesters several orders of magnitude.

We believe this work opens up many possibilities for the study of asymptotic quantum pumping and its potential applications.
Although further work is required, the proposal seems amenable to harvesting very low kinetic energy as it avoids the use of electrical rectifiers and then seems promising for powering nanoscale devices. In this context, it would be important to test the ideas proposed in more concrete examples, such as carbon nanotubes or graphene sheets under realistic conditions. One key aspect that requires a deeper study is the sensitivity of the sign of the pumped current to potential defects in the fabrication of the device. This can cause problems for parallel energy harvesting as the sign of the pumped current is not controlled externally but depends on the design of the device.

Our proposal only requires appropriate initial conditions triggered mechanically and, because of that, displacement currents should be absent. This makes asymptotic quantum pumping attractive as an alternative way of experimentally studying  quantum pumping.
Although it was not the original idea, it would also be interesting to study asymptotic quantum pumping as a thermal machine. For example, one can assume that the tip is excited by thermal noise and there is a temperature difference between the tip and the rest of the system. Appropriate working conditions should be found in this case but the idea seems appealing.

\section{Acknowledgments}

The author acknowledges useful comments and discussions with L. H. Ingaramo, L. J. Fern\'andez-Al\'azar, L. E. F. Foa Torres, and H. M. Pastawski.
This work was supported by Consejo Nacional de Investigaciones Cient\'ificas y T\'ecnicas (CONICET), Argentina; Secretar\'ia de Ciencia y Tecnolog\'ia, Universidad Nacional de C\'ordoba (SECYT-UNC), Argentina; and Ministerio de Ciencia y Tecnolog\'ia de la Provincia de C\'ordoba (MinCyT-Cor), Argentina.

\appendix
\section{Derivation of Eq. \ref{eq_dSq=dS}}
\label{app_A}

The elements of the scattering matrix of a problem can be evaluated from the Green's function of the system by using the Fisher and Lee formula \cite{Fisher1981,RevMex,cattena2014,PRBLucas} which can be written as~\cite{bode2012}
\begin{equation}
\boldsymbol{S}=\boldsymbol{I} - 2 i \boldsymbol{W}^\dagger \boldsymbol{G}^R \boldsymbol{W}. \label{eq_FisherLee}
\end{equation}
Here, $\boldsymbol{G}^R$ is the retarded Green's function
\begin{equation}
\boldsymbol{G}^R=\lim_{\eta\rightarrow 0^{+}}[(\varepsilon+i \eta) \boldsymbol{I} - \boldsymbol{H} - \boldsymbol{\Sigma}]^{-1} \label{eq_Gr}.
\end{equation}
where $\boldsymbol{H}$ is the Hamiltonian of the system without the leads, $\boldsymbol{\Sigma}$ is the self-energy due to the leads, and $\varepsilon$ is the energy of the electrons.
The matrix $\boldsymbol{W}$ comes from
\begin{equation}
\boldsymbol{\Gamma}_\alpha=\boldsymbol{W}^\dagger \boldsymbol{\Pi}_\alpha \boldsymbol{W} .
\end{equation}
where $\boldsymbol{\Pi}_\alpha$ is the projection operator onto the channel $\alpha$ of some reservoir $r$ and $\boldsymbol{\Gamma}_\alpha$ is the contributions, due to the channel $\alpha$, to the imaginary part of the self-energy $\boldsymbol{\Sigma}$, i.e. $\boldsymbol{\Gamma}=\mathrm{Im}(\boldsymbol{\Sigma})$ and $\boldsymbol{\Gamma}=\sum_\alpha \boldsymbol{\Gamma}_\alpha$.
\footnote{
A reservoir $r$ can always be described by a series of independent conduction channels $\alpha$.~\cite{RevMex} This type of description implies that the ($\boldsymbol{\Gamma}_\alpha$)s are diagonals in a tight binding representation of the system.~\cite{RevMex,cattena2014,PRBLucas} Then, the elements of $\boldsymbol{S}$, see Eq. \ref{eq_FisherLee}, result in the more familiar expression~\cite{Fisher1981,RevMex,cattena2014,PRBLucas}
$S_{\alpha \beta}=\delta_{\alpha\beta}  - 2 i \sqrt{\Gamma_{\alpha i}} G^R_{i j}\sqrt{\Gamma_{j \beta}}$
}

The derivative of $\boldsymbol{S}$ with respect to a coordinate $q_i$, that does not affect the couplings to the leads, can be written as
\begin{equation}
\frac{\partial \boldsymbol{S}}{\partial q_i}=-2 i \boldsymbol{W}^\dagger \boldsymbol{G}^R \frac{\partial \boldsymbol{H}}{\partial q_i} \boldsymbol{G}^R \boldsymbol{W}.
\end{equation}
Now, due to the particular choice of $\boldsymbol{H}$, $\boldsymbol{H}_{q_i}$, and $q_0$ ($q_0=0$) it is clear that $\frac{\partial \boldsymbol{H}}{\partial q_i}=\frac{\partial \boldsymbol{H}_{q_i}}{\partial q_i}$ and $\boldsymbol{H}(q_0)=\boldsymbol{H}_{q_i}(q_0)$,
which immediately implies Eq. \ref{eq_dSq=dS}.

\section{Derivation of Eq. \ref{eq_dS}}
\label{app_B}

To find the scattering matrix of the Hamiltonian $\hat{H}_{q_i}$ we start by linearizing it for momenta close to $\hbar k_i=\pm \hbar \pi \omega_i/L$ where $\hbar$ is the Planck constant divided by $2 \pi$.
The resulting Hamiltonian, given in terms of the counterpropagating linear channels and measuring momenta and energies from $\hbar k_i$ and  $\hbar^2 k_i^2 / ( 2 m_e)$ respectively, can be written as
\begin{equation}
\hat{H}_{q_i}= v_F \hat{p} \boldsymbol{\sigma}_z + \frac{U_0 q_i(t)}{2} \boldsymbol{\sigma}_y \Theta(x) \Theta(L-x), \label{eq_linHqi}
\end{equation}
where $\boldsymbol{\sigma}_i$ denotes the Pauli matrices in the space of the counterpropagating channels, $v_F$ is the Fermi velocity, and we do not include the electron spin for simplicity. 
The transfer matrix $\boldsymbol{M}$ of a one dimensional problem can be defined by its effect on the in- and outgoing waves
($i$ and $o$ respectively) as $(i_R,o_R)^T=\boldsymbol{M}(o_L,i_L)^T$, where  $L$ and $R$ stand for left and right leads here (not to be confused with the length of the system in Eq. \ref{eq_linHqi}). 
Then, neglecting the reflections at the boundary of the system (small $U_0$)~\cite{PRBLucas} and assuming the wave function inside of it is of the form $e^{ikx}$, one can write the transfer matrix as $M \approx e^{i L \hat k}$, which combined with Eq. \ref{eq_linHqi} yields
\begin{equation}
\boldsymbol{M}_{q_i} = \mathrm{exp}\left ( \frac{i L}{\hbar v_F}
\left [ \delta \varepsilon_i-\frac{U_0 q_i(t)}{2} \boldsymbol{\sigma}_y \right ] \boldsymbol{\sigma}_z \right )
\end{equation}
where $\delta \varepsilon_i = \left ( \varepsilon - \frac{\hbar^2 k_i^2}{2 m_e} \right )$. This equation can be rewritten as
\begin{equation}
\boldsymbol{M}_{q_i} = e^{i \lambda_L \vec{\boldsymbol{\sigma}}_{eff}}
= \boldsymbol{I} \cos \lambda_L + i \vec{\boldsymbol{\sigma}}_{eff} \sin \lambda_L
\end{equation}
where 
\begin{eqnarray}
\lambda_L = ( L / \hbar v_f) \sqrt{(\delta\varepsilon_i)^2 - (U_0 q_i / 2)^2} \notag \\
\vec{\boldsymbol{\sigma}}_{eff}=
\frac{\left [ -i (U_0 q_i /2) \boldsymbol{\sigma}_x+\delta \varepsilon_i \boldsymbol{\sigma}_z \right ]} 
{\sqrt{ (\delta \varepsilon_i )^2  -(U_0 q_i /2)^2}} .
\end{eqnarray}
The relation between $\boldsymbol{M}$ and $\boldsymbol{S}$ can be obtained from their definitions [ 
$(i_R,o_R)^T=\boldsymbol{M}(o_L,i_L)^T$ and $(o_L,o_R)^T=\boldsymbol{S}(i_L,i_R)^T$ ]. The result is
\begin{equation}
\boldsymbol{S}_{q_i} = \left( \begin{array}{cc}
    - \frac{\sin \lambda_L (U_0 q_i/ 2)}{M_{11}\sqrt{(\delta\varepsilon_i)^2 - (U_0 q_i / 2)^2}} &
     \frac{1}{M_{11}} \\
     \frac{1}{M_{11}} &
    \frac{\sin \lambda_L (U_0 q_i / 2)}{M_{11}\sqrt{(\delta\varepsilon_i)^2 - (U_0 q_i / 2)^2}}
   \end{array} \right) \label{eq_Sqi}
\end{equation}   
where
\begin{equation}
M_{11}=\cos \lambda_L - i \left( \delta \varepsilon_i / \sqrt{(\delta\varepsilon_i)^2 - (U_0 q_i / 2)^2} \right) \sin \lambda_L , 
\end{equation}
Taking the derivative of Eq. \ref{eq_Sqi} for $q_i=0$ and considering Eq. \ref{eq_dSq=dS}, gives Eq. \ref{eq_dS}.

\bibliography{./AQP-05p4-arxiv-02}

\begin{thebibliography}{38}%
\makeatletter
\providecommand \@ifxundefined [1]{%
 \@ifx{#1\undefined}
}%
\providecommand \@ifnum [1]{%
 \ifnum #1\expandafter \@firstoftwo
 \else \expandafter \@secondoftwo
 \fi
}%
\providecommand \@ifx [1]{%
 \ifx #1\expandafter \@firstoftwo
 \else \expandafter \@secondoftwo
 \fi
}%
\providecommand \natexlab [1]{#1}%
\providecommand \enquote  [1]{``#1''}%
\providecommand \bibnamefont  [1]{#1}%
\providecommand \bibfnamefont [1]{#1}%
\providecommand \citenamefont [1]{#1}%
\providecommand \href@noop [0]{\@secondoftwo}%
\providecommand \href [0]{\begingroup \@sanitize@url \@href}%
\providecommand \@href[1]{\@@startlink{#1}\@@href}%
\providecommand \@@href[1]{\endgroup#1\@@endlink}%
\providecommand \@sanitize@url [0]{\catcode `\\12\catcode `\$12\catcode
  `\&12\catcode `\#12\catcode `\^12\catcode `\_12\catcode `\%12\relax}%
\providecommand \@@startlink[1]{}%
\providecommand \@@endlink[0]{}%
\providecommand \url  [0]{\begingroup\@sanitize@url \@url }%
\providecommand \@url [1]{\endgroup\@href {#1}{\urlprefix }}%
\providecommand \urlprefix  [0]{URL }%
\providecommand \Eprint [0]{\href }%
\providecommand \doibase [0]{http://dx.doi.org/}%
\providecommand \selectlanguage [0]{\@gobble}%
\providecommand \bibinfo  [0]{\@secondoftwo}%
\providecommand \bibfield  [0]{\@secondoftwo}%
\providecommand \translation [1]{[#1]}%
\providecommand \BibitemOpen [0]{}%
\providecommand \bibitemStop [0]{}%
\providecommand \bibitemNoStop [0]{.\EOS\space}%
\providecommand \EOS [0]{\spacefactor3000\relax}%
\providecommand \BibitemShut  [1]{\csname bibitem#1\endcsname}%
\let\auto@bib@innerbib\@empty
\bibitem [{\citenamefont {Qi}\ and\ \citenamefont {McAlpine}(2010)}]{bioApp}%
  \BibitemOpen
  \bibfield  {author} {\bibinfo {author} {\bibfnamefont {Y.}~\bibnamefont
  {Qi}}\ and\ \bibinfo {author} {\bibfnamefont {M.~C.}\ \bibnamefont
  {McAlpine}},\ }\href@noop {} {\bibfield  {journal} {\bibinfo  {journal}
  {Energy Environ. Sci.}\ }\textbf {\bibinfo {volume} {3}},\ \bibinfo {pages}
  {1275} (\bibinfo {year} {2010})}\BibitemShut {NoStop}%
\bibitem [{\citenamefont {Math{\'u}na}\ \emph {et~al.}(2008)\citenamefont
  {Math{\'u}na}, \citenamefont {O’Donnell}, \citenamefont {Martinez-Catala},
  \citenamefont {Rohan},\ and\ \citenamefont {O’Flynn}}]{WLesssensors}%
  \BibitemOpen
  \bibfield  {author} {\bibinfo {author} {\bibfnamefont {C.~{\'O}.}\
  \bibnamefont {Math{\'u}na}}, \bibinfo {author} {\bibfnamefont
  {T.}~\bibnamefont {O’Donnell}}, \bibinfo {author} {\bibfnamefont {R.~V.}\
  \bibnamefont {Martinez-Catala}}, \bibinfo {author} {\bibfnamefont
  {J.}~\bibnamefont {Rohan}}, \ and\ \bibinfo {author} {\bibfnamefont
  {B.}~\bibnamefont {O’Flynn}},\ }\href@noop {} {\bibfield  {journal}
  {\bibinfo  {journal} {Talanta}\ }\textbf {\bibinfo {volume} {75}},\ \bibinfo
  {pages} {613} (\bibinfo {year} {2008})}\BibitemShut {NoStop}%
\bibitem [{\citenamefont {Balestra}(2014)}]{revBook}%
  \BibitemOpen
  \bibfield  {author} {\bibinfo {author} {\bibfnamefont {F.}~\bibnamefont
  {Balestra}},\ }\href@noop {} {\emph {\bibinfo {title} {Beyond CMOS
  Nanodevices 1, Chap. 6 Vibrational Energy Harvesting}}},\ Beyond CMOS
  Nanodevices\ (\bibinfo  {publisher} {Wiley, New York},\ \bibinfo {year}
  {2014})\BibitemShut {NoStop}%
\bibitem [{\citenamefont {Harne}\ and\ \citenamefont
  {Wang}(2013)}]{revBistable}%
  \BibitemOpen
  \bibfield  {author} {\bibinfo {author} {\bibfnamefont {R.~L.}\ \bibnamefont
  {Harne}}\ and\ \bibinfo {author} {\bibfnamefont {K.~W.}\ \bibnamefont
  {Wang}},\ }\href@noop {} {\bibfield  {journal} {\bibinfo  {journal} {Smart
  Mater. Struct.}\ }\textbf {\bibinfo {volume} {22}},\ \bibinfo {pages}
  {023001} (\bibinfo {year} {2013})}\BibitemShut {NoStop}%
\bibitem [{\citenamefont {Anton}\ and\ \citenamefont
  {Sodano}(2007)}]{revPiezo}%
  \BibitemOpen
  \bibfield  {author} {\bibinfo {author} {\bibfnamefont {S.~R.}\ \bibnamefont
  {Anton}}\ and\ \bibinfo {author} {\bibfnamefont {H.~A.}\ \bibnamefont
  {Sodano}},\ }\href@noop {} {\bibfield  {journal} {\bibinfo  {journal} {Smart
  Mater. Struct.}\ }\textbf {\bibinfo {volume} {16}},\ \bibinfo {pages} {R1}
  (\bibinfo {year} {2007})}\BibitemShut {NoStop}%
\bibitem [{\citenamefont {Cottone}\ \emph {et~al.}(2009)\citenamefont
  {Cottone}, \citenamefont {Vocca},\ and\ \citenamefont
  {Gammaitoni}}]{PRLnonlinear}%
  \BibitemOpen
  \bibfield  {author} {\bibinfo {author} {\bibfnamefont {F.}~\bibnamefont
  {Cottone}}, \bibinfo {author} {\bibfnamefont {H.}~\bibnamefont {Vocca}}, \
  and\ \bibinfo {author} {\bibfnamefont {L.}~\bibnamefont {Gammaitoni}},\
  }\href@noop {} {\bibfield  {journal} {\bibinfo  {journal} {Phys. Rev. Lett.}\
  }\textbf {\bibinfo {volume} {102}},\ \bibinfo {pages} {080601} (\bibinfo
  {year} {2009})}\BibitemShut {NoStop}%
\bibitem [{\citenamefont {Wen}\ \emph {et~al.}(2014)\citenamefont {Wen},
  \citenamefont {Yang},\ and\ \citenamefont {Wang}}]{Triboimpact}%
  \BibitemOpen
  \bibfield  {author} {\bibinfo {author} {\bibfnamefont {X.}~\bibnamefont
  {Wen}}, \bibinfo {author} {\bibfnamefont {Q.}~\bibnamefont {Yang},
  \bibfnamefont {W.~Jing}}, \ and\ \bibinfo {author} {\bibfnamefont {Z.~L.}\
  \bibnamefont {Wang}},\ }\href@noop {} {\bibfield  {journal} {\bibinfo
  {journal} {ACS Nano}\ }\textbf {\bibinfo {volume} {8}},\ \bibinfo {pages}
  {7405} (\bibinfo {year} {2014})}\BibitemShut {NoStop}%
\bibitem [{\citenamefont {Yang}\ \emph {et~al.}(2014)\citenamefont {Yang},
  \citenamefont {Chen}, \citenamefont {Yang}, \citenamefont {Zhang},
  \citenamefont {Yang}, \citenamefont {Bai},\ and\ \citenamefont
  {Wang}}]{TriboBroad}%
  \BibitemOpen
  \bibfield  {author} {\bibinfo {author} {\bibfnamefont {J.}~\bibnamefont
  {Yang}}, \bibinfo {author} {\bibfnamefont {J.}~\bibnamefont {Chen}}, \bibinfo
  {author} {\bibfnamefont {Y.}~\bibnamefont {Yang}}, \bibinfo {author}
  {\bibfnamefont {H.}~\bibnamefont {Zhang}}, \bibinfo {author} {\bibfnamefont
  {W.}~\bibnamefont {Yang}}, \bibinfo {author} {\bibfnamefont {Y.}~\bibnamefont
  {Bai}, \bibfnamefont {P.~Su}}, \ and\ \bibinfo {author} {\bibfnamefont
  {Z.~L.}\ \bibnamefont {Wang}},\ }\href@noop {} {\bibfield  {journal}
  {\bibinfo  {journal} {Adv. Energy Mater.}\ }\textbf {\bibinfo {volume} {4}},\
  \bibinfo {pages} {1301322} (\bibinfo {year} {2014})}\BibitemShut {NoStop}%
\bibitem [{\citenamefont {Kim}\ \emph {et~al.}(2013)\citenamefont {Kim},
  \citenamefont {Prada}, \citenamefont {Platero},\ and\ \citenamefont
  {Blick}}]{PRLDdots}%
  \BibitemOpen
  \bibfield  {author} {\bibinfo {author} {\bibfnamefont {C.}~\bibnamefont
  {Kim}}, \bibinfo {author} {\bibfnamefont {M.}~\bibnamefont {Prada}}, \bibinfo
  {author} {\bibfnamefont {G.}~\bibnamefont {Platero}}, \ and\ \bibinfo
  {author} {\bibfnamefont {R.~H.}\ \bibnamefont {Blick}},\ }\href@noop {}
  {\bibfield  {journal} {\bibinfo  {journal} {Phys. Rev. Lett.}\ }\textbf
  {\bibinfo {volume} {111}},\ \bibinfo {pages} {197202} (\bibinfo {year}
  {2013})}\BibitemShut {NoStop}%
\bibitem [{\citenamefont {Hartmann}\ \emph {et~al.}(2015)\citenamefont
  {Hartmann}, \citenamefont {Pfeffer}, \citenamefont {H{\"o}fling},
  \citenamefont {Kamp},\ and\ \citenamefont {Worschech}}]{PRLnoiseRect}%
  \BibitemOpen
  \bibfield  {author} {\bibinfo {author} {\bibfnamefont {F.}~\bibnamefont
  {Hartmann}}, \bibinfo {author} {\bibfnamefont {P.}~\bibnamefont {Pfeffer}},
  \bibinfo {author} {\bibfnamefont {S.}~\bibnamefont {H{\"o}fling}}, \bibinfo
  {author} {\bibfnamefont {M.}~\bibnamefont {Kamp}}, \ and\ \bibinfo {author}
  {\bibfnamefont {L.}~\bibnamefont {Worschech}},\ }\href@noop {} {\bibfield
  {journal} {\bibinfo  {journal} {Phys. Rev. Lett.}\ }\textbf {\bibinfo
  {volume} {114}},\ \bibinfo {pages} {146805} (\bibinfo {year}
  {2015})}\BibitemShut {NoStop}%
\bibitem [{\citenamefont {Wang}(2008)}]{wang2008}%
  \BibitemOpen
  \bibfield  {author} {\bibinfo {author} {\bibfnamefont {Z.~L.}\ \bibnamefont
  {Wang}},\ }\href@noop {} {\bibfield  {journal} {\bibinfo  {journal} {Advanced
  Functional Materials}\ }\textbf {\bibinfo {volume} {18}},\ \bibinfo {pages}
  {3553} (\bibinfo {year} {2008})}\BibitemShut {NoStop}%
\bibitem [{\citenamefont {Xu}\ \emph {et~al.}(2010)\citenamefont {Xu},
  \citenamefont {H.},\ and\ \citenamefont {Wang}}]{sheng2010}%
  \BibitemOpen
  \bibfield  {author} {\bibinfo {author} {\bibfnamefont {S.}~\bibnamefont
  {Xu}}, \bibinfo {author} {\bibfnamefont {B.~J.}\ \bibnamefont {H.}}, \ and\
  \bibinfo {author} {\bibfnamefont {Z.~L.}\ \bibnamefont {Wang}},\ }\href@noop
  {} {\bibfield  {journal} {\bibinfo  {journal} {Nature Communications}\
  }\textbf {\bibinfo {volume} {1}},\ \bibinfo {pages} {93} (\bibinfo {year}
  {2010})}\BibitemShut {NoStop}%
\bibitem [{\citenamefont {B{\"u}ttiker}\ \emph {et~al.}(1994)\citenamefont
  {B{\"u}ttiker}, \citenamefont {Thomas},\ and\ \citenamefont
  {Pr{\^e}tre}}]{ZPBButtiker}%
  \BibitemOpen
  \bibfield  {author} {\bibinfo {author} {\bibfnamefont {M.}~\bibnamefont
  {B{\"u}ttiker}}, \bibinfo {author} {\bibfnamefont {H.}~\bibnamefont
  {Thomas}}, \ and\ \bibinfo {author} {\bibfnamefont {A.}~\bibnamefont
  {Pr{\^e}tre}},\ }\href@noop {} {\bibfield  {journal} {\bibinfo  {journal} {Z.
  Phys. B}\ }\textbf {\bibinfo {volume} {94}},\ \bibinfo {pages} {133}
  (\bibinfo {year} {1994})}\BibitemShut {NoStop}%
\bibitem [{\citenamefont {Brouwer}(1998)}]{Brouwer}%
  \BibitemOpen
  \bibfield  {author} {\bibinfo {author} {\bibfnamefont {P.~W.}\ \bibnamefont
  {Brouwer}},\ }\href@noop {} {\bibfield  {journal} {\bibinfo  {journal} {Phys.
  Rev. B}\ }\textbf {\bibinfo {volume} {58}},\ \bibinfo {pages} {R10 135}
  (\bibinfo {year} {1998})}\BibitemShut {NoStop}%
\bibitem [{\citenamefont {Avron}\ \emph {et~al.}(2000)\citenamefont {Avron},
  \citenamefont {Elgart}, \citenamefont {Graf},\ and\ \citenamefont
  {Sadun}}]{avron2000}%
  \BibitemOpen
  \bibfield  {author} {\bibinfo {author} {\bibfnamefont {J.~E.}\ \bibnamefont
  {Avron}}, \bibinfo {author} {\bibfnamefont {A.}~\bibnamefont {Elgart}},
  \bibinfo {author} {\bibfnamefont {G.~M.}\ \bibnamefont {Graf}}, \ and\
  \bibinfo {author} {\bibfnamefont {L.}~\bibnamefont {Sadun}},\ }\href@noop {}
  {\bibfield  {journal} {\bibinfo  {journal} {Phys. Rev. B}\ }\textbf {\bibinfo
  {volume} {62}},\ \bibinfo {pages} {R10618} (\bibinfo {year}
  {2000})}\BibitemShut {NoStop}%
\bibitem [{\citenamefont {Watson}\ \emph {et~al.}(2003)\citenamefont {Watson},
  \citenamefont {Potok}, \citenamefont {Marcus},\ and\ \citenamefont
  {Umansky}}]{watson2003}%
  \BibitemOpen
  \bibfield  {author} {\bibinfo {author} {\bibfnamefont {S.~K.}\ \bibnamefont
  {Watson}}, \bibinfo {author} {\bibfnamefont {R.~M.}\ \bibnamefont {Potok}},
  \bibinfo {author} {\bibfnamefont {C.~M.}\ \bibnamefont {Marcus}}, \ and\
  \bibinfo {author} {\bibfnamefont {V.}~\bibnamefont {Umansky}},\ }\href@noop
  {} {\bibfield  {journal} {\bibinfo  {journal} {Phys. Rev. Lett.}\ }\textbf
  {\bibinfo {volume} {91}},\ \bibinfo {pages} {258301} (\bibinfo {year}
  {2003})}\BibitemShut {NoStop}%
\bibitem [{\citenamefont {Foa~Torres}(2005)}]{foa2005}%
  \BibitemOpen
  \bibfield  {author} {\bibinfo {author} {\bibfnamefont {L.~E.~F.}\
  \bibnamefont {Foa~Torres}},\ }\href@noop {} {\bibfield  {journal} {\bibinfo
  {journal} {Phys. Rev. B}\ }\textbf {\bibinfo {volume} {72}},\ \bibinfo
  {pages} {245339} (\bibinfo {year} {2005})}\BibitemShut {NoStop}%
\bibitem [{\citenamefont {Strass}\ \emph {et~al.}(2005)\citenamefont {Strass},
  \citenamefont {H\"anggi},\ and\ \citenamefont {Kohler}}]{strass2005}%
  \BibitemOpen
  \bibfield  {author} {\bibinfo {author} {\bibfnamefont {M.}~\bibnamefont
  {Strass}}, \bibinfo {author} {\bibfnamefont {P.}~\bibnamefont {H\"anggi}}, \
  and\ \bibinfo {author} {\bibfnamefont {S.}~\bibnamefont {Kohler}},\
  }\href@noop {} {\bibfield  {journal} {\bibinfo  {journal} {Phys. Rev. Lett.}\
  }\textbf {\bibinfo {volume} {95}},\ \bibinfo {pages} {130601} (\bibinfo
  {year} {2005})}\BibitemShut {NoStop}%
\bibitem [{\citenamefont {Splettstoesser}\ \emph {et~al.}(2005)\citenamefont
  {Splettstoesser}, \citenamefont {Governale}, \citenamefont {K\"onig},\ and\
  \citenamefont {Fazio}}]{splettstoesser2005}%
  \BibitemOpen
  \bibfield  {author} {\bibinfo {author} {\bibfnamefont {J.}~\bibnamefont
  {Splettstoesser}}, \bibinfo {author} {\bibfnamefont {M.}~\bibnamefont
  {Governale}}, \bibinfo {author} {\bibfnamefont {J.}~\bibnamefont {K\"onig}},
  \ and\ \bibinfo {author} {\bibfnamefont {R.}~\bibnamefont {Fazio}},\
  }\href@noop {} {\bibfield  {journal} {\bibinfo  {journal} {Phys. Rev. Lett.}\
  }\textbf {\bibinfo {volume} {95}},\ \bibinfo {pages} {246803} (\bibinfo
  {year} {2005})}\BibitemShut {NoStop}%
\bibitem [{\citenamefont {Arrachea}\ and\ \citenamefont
  {Moskalets}(2006)}]{arrachea2006}%
  \BibitemOpen
  \bibfield  {author} {\bibinfo {author} {\bibfnamefont {L.}~\bibnamefont
  {Arrachea}}\ and\ \bibinfo {author} {\bibfnamefont {M.}~\bibnamefont
  {Moskalets}},\ }\href@noop {} {\bibfield  {journal} {\bibinfo  {journal}
  {Phys. Rev. B}\ }\textbf {\bibinfo {volume} {74}},\ \bibinfo {pages} {245322}
  (\bibinfo {year} {2006})}\BibitemShut {NoStop}%
\bibitem [{\citenamefont {Nakajima}\ \emph {et~al.}(2016)\citenamefont
  {Nakajima}, \citenamefont {Tomita}, \citenamefont {Taie}, \citenamefont
  {Ichinose}, \citenamefont {Ozawa}, \citenamefont {Wang}, \citenamefont
  {Troyer},\ and\ \citenamefont {Takahashi}}]{nakajima2016}%
  \BibitemOpen
  \bibfield  {author} {\bibinfo {author} {\bibfnamefont {S.}~\bibnamefont
  {Nakajima}}, \bibinfo {author} {\bibfnamefont {T.}~\bibnamefont {Tomita}},
  \bibinfo {author} {\bibfnamefont {S.}~\bibnamefont {Taie}}, \bibinfo {author}
  {\bibfnamefont {T.}~\bibnamefont {Ichinose}}, \bibinfo {author}
  {\bibfnamefont {H.}~\bibnamefont {Ozawa}}, \bibinfo {author} {\bibfnamefont
  {L.}~\bibnamefont {Wang}}, \bibinfo {author} {\bibfnamefont {M.}~\bibnamefont
  {Troyer}}, \ and\ \bibinfo {author} {\bibfnamefont {Y.}~\bibnamefont
  {Takahashi}},\ }\href@noop {} {\bibfield  {journal} {\bibinfo  {journal}
  {Nat. Phys.}\ }\textbf {\bibinfo {volume} {12}},\ \bibinfo {pages} {296}
  (\bibinfo {year} {2016})}\BibitemShut {NoStop}%
\bibitem [{\citenamefont {Schweizer}\ \emph {et~al.}(2016)\citenamefont
  {Schweizer}, \citenamefont {Lohse}, \citenamefont {Citro},\ and\
  \citenamefont {Bloch}}]{schweizer2016}%
  \BibitemOpen
  \bibfield  {author} {\bibinfo {author} {\bibfnamefont {C.}~\bibnamefont
  {Schweizer}}, \bibinfo {author} {\bibfnamefont {M.}~\bibnamefont {Lohse}},
  \bibinfo {author} {\bibfnamefont {R.}~\bibnamefont {Citro}}, \ and\ \bibinfo
  {author} {\bibfnamefont {I.}~\bibnamefont {Bloch}},\ }\href@noop {}
  {\bibfield  {journal} {\bibinfo  {journal} {Phys. Rev. Lett.}\ }\textbf
  {\bibinfo {volume} {117}},\ \bibinfo {pages} {170405} (\bibinfo {year}
  {2016})}\BibitemShut {NoStop}%
\bibitem [{Note1()}]{Note1}%
  \BibitemOpen
  \bibinfo {note} {Interesting physical systems where this condition can be
  found are carbon nanotubes and graphene sheets for example~\cite
  {foatorres2014}}\BibitemShut {NoStop}%
\bibitem [{Note2()}]{Note2}%
  \BibitemOpen
  \bibinfo {note} {The low-temperature limit of the emissivity is used just for
  simplicity. For finite temperatures an extra integral should be added to the
  formulas of the scattering factor as now \begin {equation} \protect \frac
  {dn_r}{dq_i} = - \DOTSI \intop \ilimits@ \protect \frac {df}{d\varepsilon
  }{\DOTSB \sum@ \slimits@ _{\beta , \alpha \in r} \protect \frac {1}{2 \pi }
  \protect \mathrm {Im} \left [ \protect \frac {\partial S_{\alpha \beta
  }}{\partial q_i} S^{\ast }_{\alpha \beta } \right ]} d\varepsilon , \end
  {equation} where $f$ is the Fermi function}\BibitemShut {NoStop}%
\bibitem [{Note3()}]{Note3}%
  \BibitemOpen
  \bibinfo {note} {To see that, take the integral $\DOTSI \intop \ilimits@
  _0^{\infty } q_i \protect \mathaccentV {dot}05F{q}_j dt$ and divide it into
  time intervals that correspond to the different closed trajectories, $\DOTSI
  \intop \ilimits@ _0^{\infty } dt = \DOTSI \intop \ilimits@ _0^{t_1}
  dt+...+\DOTSI \intop \ilimits@ _{t_i}^{t_{i+1}} dt+...$. Then simply change
  the variables of the integrals as $\DOTSI \intop \ilimits@ _{t_i}^{t_{i+1}}
  q_i\protect \mathaccentV {dot}05F{q}_j d t = \DOTSI \ointop \ilimits@
  q_i(q_j) d q_j$. The last integral is the area enclosed by the particular
  segment $(i,i+1)$ of the total trajectory}\BibitemShut {NoStop}%
\bibitem [{Note4()}]{Note4}%
  \BibitemOpen
  \bibinfo {note} {\label {note1}Displacement currents arise from the
  capacitive coupling of time-dependent gate voltages with the reservoirs.
  These currents typically hinder the detection of pumping currents.\cite
  {brouwer2001}. In our case, gate voltages are not necessary in general but
  even in the case of using them, as may be the case for proposals similar to
  those shown in Fig. \ref {fig_scheme}, they are time independent. The time
  dependence is in the deformation of the system itself, which is independent
  of any external agent.}\BibitemShut {Stop}%
\bibitem [{Note5()}]{Note5}%
  \BibitemOpen
  \bibinfo {note} {An electret is a dielectric material that has a
  quasi-permanent electric charge or dipolar polarization. An example of its
  application for energy harvesting can be found in Ref. ~\protect
  \rev@citealpnum {electret}.}\BibitemShut {Stop}%
\bibitem [{\citenamefont {Fisher}\ and\ \citenamefont
  {Lee}(1981)}]{Fisher1981}%
  \BibitemOpen
  \bibfield  {author} {\bibinfo {author} {\bibfnamefont {D.~S.}\ \bibnamefont
  {Fisher}}\ and\ \bibinfo {author} {\bibfnamefont {P.~A.}\ \bibnamefont
  {Lee}},\ }\href@noop {} {\bibfield  {journal} {\bibinfo  {journal} {Phys.
  Rev. B}\ }\textbf {\bibinfo {volume} {23}},\ \bibinfo {pages} {6851}
  (\bibinfo {year} {1981})}\BibitemShut {NoStop}%
\bibitem [{\citenamefont {Bustos-Mar{\'u}n}\ \emph {et~al.}(2013)\citenamefont
  {Bustos-Mar{\'u}n}, \citenamefont {Refael},\ and\ \citenamefont {von
  Oppen}}]{PRLBustos}%
  \BibitemOpen
  \bibfield  {author} {\bibinfo {author} {\bibfnamefont {R.~A.}\ \bibnamefont
  {Bustos-Mar{\'u}n}}, \bibinfo {author} {\bibfnamefont {G.}~\bibnamefont
  {Refael}}, \ and\ \bibinfo {author} {\bibfnamefont {F.}~\bibnamefont {von
  Oppen}},\ }\href@noop {} {\bibfield  {journal} {\bibinfo  {journal} {Phys.
  Rev. Lett.}\ }\textbf {\bibinfo {volume} {111}},\ \bibinfo {pages} {060802}
  (\bibinfo {year} {2013})}\BibitemShut {NoStop}%
\bibitem [{\citenamefont {Fern{\'a}ndez-Alc{\'a}zar}\ \emph
  {et~al.}(2017)\citenamefont {Fern{\'a}ndez-Alc{\'a}zar}, \citenamefont
  {Pastawski},\ and\ \citenamefont {Bustos-Mar{\'u}n}}]{PRBLucas}%
  \BibitemOpen
  \bibfield  {author} {\bibinfo {author} {\bibfnamefont {L.~J.}\ \bibnamefont
  {Fern{\'a}ndez-Alc{\'a}zar}}, \bibinfo {author} {\bibfnamefont {H.~M.}\
  \bibnamefont {Pastawski}}, \ and\ \bibinfo {author} {\bibfnamefont {R.~A.}\
  \bibnamefont {Bustos-Mar{\'u}n}},\ }\href@noop {} {\bibfield  {journal}
  {\bibinfo  {journal} {Phys. Rev. B}\ }\textbf {\bibinfo {volume} {95}},\
  \bibinfo {pages} {155410} (\bibinfo {year} {2017})}\BibitemShut {NoStop}%
\bibitem [{\citenamefont {{Pastawski}}\ and\ \citenamefont
  {{Medina}}(2001)}]{RevMex}%
  \BibitemOpen
  \bibfield  {author} {\bibinfo {author} {\bibfnamefont {H.~M.}\ \bibnamefont
  {{Pastawski}}}\ and\ \bibinfo {author} {\bibfnamefont {E.}~\bibnamefont
  {{Medina}}},\ }\href@noop {} {\bibfield  {journal} {\bibinfo  {journal} {Rev.
  Mex. Fis.}\ }\textbf {\bibinfo {volume} {47S1}},\ \bibinfo {pages} {1}
  (\bibinfo {year} {2001})}\BibitemShut {NoStop}%
\bibitem [{\citenamefont {Cattena}\ \emph {et~al.}(2014)\citenamefont
  {Cattena}, \citenamefont {Fern\'andez-Alc\'azar}, \citenamefont
  {Bustos-Mar\'un}, \citenamefont {Nozaki},\ and\ \citenamefont
  {Pastawski}}]{cattena2014}%
  \BibitemOpen
  \bibfield  {author} {\bibinfo {author} {\bibfnamefont {C.~J.}\ \bibnamefont
  {Cattena}}, \bibinfo {author} {\bibfnamefont {L.~J.}\ \bibnamefont
  {Fern\'andez-Alc\'azar}}, \bibinfo {author} {\bibfnamefont {R.~A.}\
  \bibnamefont {Bustos-Mar\'un}}, \bibinfo {author} {\bibfnamefont
  {D.}~\bibnamefont {Nozaki}}, \ and\ \bibinfo {author} {\bibfnamefont {H.~M.}\
  \bibnamefont {Pastawski}},\ }\href@noop {} {\bibfield  {journal} {\bibinfo
  {journal} {Journal of Physics: Condensed Matter}\ }\textbf {\bibinfo {volume}
  {26}},\ \bibinfo {pages} {345304} (\bibinfo {year} {2014})}\BibitemShut
  {NoStop}%
\bibitem [{Note6()}]{Note6}%
  \BibitemOpen
  \bibinfo {note} {To recover the expression for $Q_L$ shown in Eq. \ref
  {eq_Qtotal} we assumed that there is not an accumulation of charges in the
  system $Q_L=-Q_R$ and that the equilibrium positions do not change ($q_i(t_0)
  \approx q_i(t_\infty )$). The latter is reasonable for $Q_L$ much smaller
  than the total charge accumulated in the capacitor, which implies $V(t_0)
  \approx V(t_\infty )$}\BibitemShut {NoStop}%
\bibitem [{\citenamefont {Bode}\ \emph {et~al.}(2012)\citenamefont {Bode},
  \citenamefont {Viola~Kusminskiy}, \citenamefont {Egger},\ and\ \citenamefont
  {von Oppen}}]{bode2012}%
  \BibitemOpen
  \bibfield  {author} {\bibinfo {author} {\bibfnamefont {N.}~\bibnamefont
  {Bode}}, \bibinfo {author} {\bibfnamefont {S.}~\bibnamefont
  {Viola~Kusminskiy}}, \bibinfo {author} {\bibfnamefont {R.}~\bibnamefont
  {Egger}}, \ and\ \bibinfo {author} {\bibfnamefont {F.}~\bibnamefont {von
  Oppen}},\ }\href@noop {} {\bibfield  {journal} {\bibinfo  {journal}
  {Beilstein Journal of Nanotechnology}\ }\textbf {\bibinfo {volume} {3}},\
  \bibinfo {pages} {144} (\bibinfo {year} {2012})}\BibitemShut {NoStop}%
\bibitem [{Note7()}]{Note7}%
  \BibitemOpen
  \bibinfo {note} {A reservoir $r$ can always be described by a series of
  independent conduction channels $\alpha $.~\cite {RevMex} This type of
  description implies that the ($\protect \boldsymbol {\Gamma }_\alpha $)s are
  diagonals in a tight binding representation of the system.~\cite
  {RevMex,cattena2014,PRBLucas} Then, the elements of $\protect \boldsymbol
  {S}$, see Eq. \ref {eq_FisherLee}, result in the more familiar
  expression~\cite {Fisher1981,RevMex,cattena2014,PRBLucas} $S_{\alpha \beta
  }=\delta _{\alpha \beta } - 2 i \protect \sqrt {\Gamma _{\alpha i}} G^R_{i
  j}\protect \sqrt {\Gamma _{j \beta }}$}\BibitemShut {NoStop}%
\bibitem [{\citenamefont {Foa~Torres}\ \emph {et~al.}(2014)\citenamefont
  {Foa~Torres}, \citenamefont {Roche},\ and\ \citenamefont
  {Charlier}}]{foatorres2014}%
  \BibitemOpen
  \bibfield  {author} {\bibinfo {author} {\bibfnamefont {L.}~\bibnamefont
  {Foa~Torres}}, \bibinfo {author} {\bibfnamefont {S.}~\bibnamefont {Roche}}, \
  and\ \bibinfo {author} {\bibfnamefont {J.}~\bibnamefont {Charlier}},\
  }\href@noop {} {\emph {\bibinfo {title} {Introduction to Graphene-Based
  Nanomaterials: From Electronic Structure to Quantum Transport}}}\ (\bibinfo
  {publisher} {Cambridge University Press, Cambridge, UK},\ \bibinfo {year}
  {2014})\BibitemShut {NoStop}%
\bibitem [{\citenamefont {Brouwer}(2001)}]{brouwer2001}%
  \BibitemOpen
  \bibfield  {author} {\bibinfo {author} {\bibfnamefont {P.~W.}\ \bibnamefont
  {Brouwer}},\ }\href@noop {} {\bibfield  {journal} {\bibinfo  {journal} {Phys.
  Rev. B}\ }\textbf {\bibinfo {volume} {63}},\ \bibinfo {pages} {121303}
  (\bibinfo {year} {2001})}\BibitemShut {NoStop}%
\bibitem [{\citenamefont {Tao}\ \emph {et~al.}(2014)\citenamefont {Tao},
  \citenamefont {Liu}, \citenamefont {Lye}, \citenamefont {Miao},\ and\
  \citenamefont {Hu}}]{electret}%
  \BibitemOpen
  \bibfield  {author} {\bibinfo {author} {\bibfnamefont {K.}~\bibnamefont
  {Tao}}, \bibinfo {author} {\bibfnamefont {S.}~\bibnamefont {Liu}}, \bibinfo
  {author} {\bibfnamefont {S.~W.}\ \bibnamefont {Lye}}, \bibinfo {author}
  {\bibfnamefont {J.}~\bibnamefont {Miao}}, \ and\ \bibinfo {author}
  {\bibfnamefont {X.}~\bibnamefont {Hu}},\ }\href@noop {} {\bibfield  {journal}
  {\bibinfo  {journal} {J. Micromech. Microeng.}\ }\textbf {\bibinfo {volume}
  {24}},\ \bibinfo {pages} {065022} (\bibinfo {year} {2014})}\BibitemShut
  {NoStop}%
\end{thebibliography}%
\end{document}